\documentclass[prb,superscriptaddress,twocolumn,eqsecnum]{revtex4-1}

\usepackage{hyperref}
\usepackage{amsthm}
\usepackage{graphicx}
\usepackage{amsfonts}
\usepackage[figuresright]{rotating}  
\usepackage{amssymb}
\usepackage{amsmath}
\usepackage{psfrag}
\usepackage{subfigure}
\usepackage{multirow}
\usepackage{tabularx}
\usepackage{bm}
\usepackage{braket}

\usepackage{color}

\newtheorem*{theorem}{Theorem}

\newcommand{\bea}{\begin{eqnarray}}
\newcommand{\eea}{\end{eqnarray}}

\newcommand{\be}{\begin{equation}}
\newcommand{\ee}{\end{equation}}



\begin{document}

\title{Adiabatic continuity, wavefunction overlap and topological phase transitions}
\author{Jiahua Gu}
\author{Kai Sun}
\affiliation{Department of Physics, University of Michigan, Ann Arbor, MI 48109, USA}

\date{\today}

\begin{abstract}
In this article, we study the relation between wavefunction overlap and adiabatic continuity in gapped quantum systems.
We show that for two band insulators, a scalar function can be defined in the 
momentum space, which characterizes the wavefunction overlap between Bloch states in the two insulators. 
If this overlap is nonzero for all momentum points 
in the Brillouin zone, these two insulators are adiabatically connected, i.e. we can deform one insulator into the other 
smoothly without closing the band gap. In addition, we further prove that this adiabatic path preserves all the 
symmetries of the insulators. The existence of such an adiabatic path implies that two insulators with nonzero wavefunction 
overlap belong to the same topological phase. This relation, between adiabatic continuity and wavefunction overlap,
can be further generalized to correlated systems. The generalized relation cannot be applied to study generic 
many-body systems in the thermodynamic limit, because of the orthogonality catastrophe.
However, for certain interacting systems (e.g. quantum Hall systems), the quantum wavefunction overlap 
can be utilized to distinguish different quantum states. Experimental implications are also discussed.
\end{abstract}

\maketitle


\section{Introduction}
In quantum many-body systems, quantum phase transitions are among the most fascinating 
phenomena \cite{sachdev1999}. 
From the point of the view of adiabatic continuity, a quantum phase 
transition can be characterized by the absence of an adiabatic path between ground states of 
quantum systems. Consider two quantum many-body systems in their ground states.
If an adiabatic path can be constructed to smoothly deform one system into the other without any singularity, 
these two quantum states can be classified into the same quantum phase. On the other hand, if it is impossible 
to adiabatically deform one quantum system into the other,  without going through some singular point (or
some intermediate phase), these two quantum states belong to different quantum phases of matter 
and the singular point, which arises when we try to deform one system into the other, is a quantum phase transition point.

In general, quantum phase transitions can be largely classified into two categories, Landau-type and topological, depending
on the origin of the singularity. In the first category, the two quantum phases separated by a quantum phase transition 
have different symmetries, i.e. certain symmetry is broken spontaneously as we move across the phase boundary.
Similar to a classical (thermal) phase transition, the difference in symmetry implies that it is impossible
for these two quantum states to smoothly evolve into each other without undergoing a quantum phase transition. 
In the second category, the two quantum phases have the same symmetry, but their ground-state wavefunctions
have different topological structures. 
For a gapped quantum system, where a finite energy gap exists between the ground state and the excited ones, 
the topology of the ground state wavefunction cannot change in any adiabatic procedure without closing the 
excitation gap. Thus, if the ground state wavefunctions of two gapped quantum systems have different topology,  
as we try to deform one into the other, a singularity point must arise, at which the energy gap 
closes and the ground-state wavefunction changes its topology. This singular point is known as a 
topological phase transition. Such a topological transition can take place even in the absence of interactions, 
e.g. in non-interacting band 
insulators~\cite{haldane1988, hasan2010, qiRMP, Bernevig2013, kitaev2009art, schnyder2008, Moore2008, Fu2011}.

In this article, we study adiabatic continuity between quantum states in gapped quantum systems focusing 
on the following question: {\it  for two (arbitrary) quantum states, how can we determine whether 
a gapped adiabatic path between these two states exists or not}?
 More precisely, we want to determine,
for two quantum states $\ket{\psi_1}$ and $\ket{\psi_2}$, whether it is possible or not to construct a 
gapped Hamiltonian $H(\alpha)$, where $\alpha$ is some control parameter, such that as we tune 
the value of the control parameter $\alpha$, the ground state of the Hamiltonian changes smoothly from 
$\ket{\psi_1}$ to $\ket{\psi_2}$. It must be emphasized that here we require that the Hamiltonian remains gapped
for this adiabatic procedure, i.e., the energy gap between the ground and excited states never vanishes. 
As discussed above, the answer to this question is of direct relevance to the study of quantum phase transitions
between gapped quantum systems, including topological phase transitions.

For band-insulators, we find that regardless of the symmetry and microscopic details, as long as the Bloch wavefunctions 
(of the valence bands) of two insulators have finite wavefunction overlap, an adiabatic path can be constructed, connecting 
the two insulators without closing the insulating gap. For the study of topological band insulators, this conclusion implies
that two band insulators with finite wavefunction overlap must have the same topology, 
i.e, all topological indices take the same value in the two insulators. 
This result also implies that for two insulators with different topology, there must exist at least one momentum point
in the Brillouin zone, at which the Bloch waves in these two insulators are orthogonal to each other, i.e. 
the wavefunctions have zero overlap.

This conclusion can be easily generalized to interacting systems, i.e., if two quantum states have finite wavefunction 
overlap, regardless of microscopic details, a gapped adiabatic path can be defined to connect these two states.
However, as pointed out below, this conclusion cannot be applied to study generic quantum many-body systems 
and quantum phase transitions, due to the orthogonality catastrophe~\cite{Anderson1967}, 
which says that in the thermodynamic limit, even for two quantum states in the same quantum phase, 
the wavefunction overlap will vanish due to the infinite size of the system.  As a result, the wavefunction overlap, 
which is always zero in the thermodynamic limit, doesn't carry useful information about quantum phases and 
adiabatic continuity. This is in sharp contrast to noninteracting systems, e.g. band insulators, 
where we can utilize single-particle Bloch waves, which do not suffer from the orthogonality catastrophe.
In this article, we show that the problem caused by the orthogonality 
catastrophe can be resolved in certain interacting systems,
including integer and fractional quantum Hall systems~\cite{klitzing1980, Tsui1982}, 
and integer and fractional Chern 
insulators~\cite{haldane1988, Kol1993, Sorensen2005, Moller2009,Tang2011, Sun2011, Neupert2011, Sheng2011, Regnault2011,
Parameswaran2013},
utilizing  various schemes, e.g., by studying systems with finite size or  factorizing the many-body 
wavefunction. 

In this article, we study adiabatic continuity in band insulators in Sec.~\ref{sec:band_insulator}.
Then in Sec.~\ref{sec:interacting}, we generalize the conclusion to interacting systems.
In Sec.~\ref{sec:interaction_phase_transition}, we discuss how to utilize this result to 
study quantum phase transitions in the presence of interactions. Two examples, will be discussed.
Finally, we conclude the article by discussing possible implications 
in experimental and numerical studies. Details about the calculations and proofs  are shown in Appendix.

\section{Band insulators}
\label{sec:band_insulator}
For band insulators, if we only focus on the qualitative properties, interactions can often be ignored.
Within the non-interacting approximation, the quantum wavefunction of a band insulator is the (anti-symmetrized) 
product of Bloch-wave states. Because of the lattice translational symmetry,
Bloch states with different crystal momenta decouple from one another. 
Therefore, we can examine wavefunction overlap 
at each momentum point separately.

In this section, we focus on the non-interacting regime. First, we prove that for two band insulators,  
the wavefunction overlap between the many-body ground states factorizes into the product of (Bloch-wavefunction) overlaps 
at each momentum point. Then, we will show that if the overlap remains finite for all momenta, 
the two insulators are adiabatically connected, i.e. we can adiabatically deform the wavefunction of one insulator into the other 
without closing the insulating gap or breaking any symmetries. 

This conclusion immediately implies that (a) if two band insulators belong to two different quantum phases 
(i.e. it is impossible to deform one state into the other without closing the insulating gap), there must exist (at least) 
one momentum point $\mathbf{k}^*$, at which the (Bloch-wavefunction) overlap between the two insulators vanishes, 
and (b) if the Bloch wavefunctions of two band insulators have finite overlap at all momenta,
these two insulators must belong to the same quantum phase.

We start the discussion by considering insulators with only one valence band (Sec.~\ref{sec:one-band}). 
Then in Sec.~\ref{sec:multi_bands}, we will generalize the conclusions to generic cases with multiple valence bands.

\subsection{Insulators with one valence band}
\label{sec:one-band}
In this section, we consider two band insulators, dubbed insulator $I$ and insulator $II$, each of which has only 
one valence band.  More generic situations (with more than one valence bands) will be studied in the next section.

\subsubsection{wavefunction overlap}
Within the non-interacting approximation, the many-body ground states of these two insulators can be written as
\begin{align}
\ket{\textrm{G}_I} =\prod_{\mathbf{k}} c^\dagger_{\mathbf{k}} \ket{0}
\label{eq:G_I}
\\
\ket{\textrm{G}_{II}} =\prod_{\mathbf{k}} d^\dagger_{\mathbf{k}} \ket{0}
\label{eq:G_II}
\end{align}
where $\ket{\textrm{G}_{I}}$ and $\ket{\textrm{G}_{II}}$ are the (many-body) ground states of the two insulators respectively. 
$\ket{0}$ represents the vacuum, i.e. the quantum state with no electrons. $c^\dagger_{\mathbf{k}}$ 
($d^\dagger_{\mathbf{k}}$) is the creation operator which creates a particle in the Bloch state of the valence band 
in insulator $I$ (insulator $II$) at crystal momentum $\mathbf{k}$.  $\prod_{\mathbf{k}}$ 
represents the product over all momenta in the Brillouin zone.

It is straightforward to verify that the overlap between the two ground states factorizes as
\begin{align}
|\braket{\textrm{G}_I|\textrm{G}_{II}}|=\prod_{\mathbf{k}} |\phi(\mathbf{k})|
\label{eq:factor_overlap_one_band}
\end{align}
where $ \phi(\mathbf{k})$ is the overlap between Bloch waves at crystal momentum $\mathbf{k}$
\begin{align}
 \phi(\mathbf{k})=\braket{0| c_{\mathbf{k}}d^\dagger_{\mathbf{k}} |0}
\end{align}
In the language of first quantization, this Bloch-wave overlap is
\begin{align}
 \phi(\mathbf{k})=\braket{\psi^{I}(\mathbf{k})|\psi^{II}(\mathbf{k})}
\label{eq:overlap_one_band}
\end{align}
where 
\begin{align}
\ket{\psi^{I}(\mathbf{k})}=c^\dagger_{\mathbf{k}}\ket{0}
\label{eq:bloch_one_band_1}
\\
\ket{\psi^{II}(\mathbf{k})}=d^\dagger_{\mathbf{k}}\ket{0}
\label{eq:bloch_one_band_2}
\end{align}
are the Bloch waves of the valence bands in insulators $I$ and $II$ respectively.

\subsubsection{the adiabatic path between two insulators}

Define a new Bloch state
\begin{align}
\ket{\Psi(\mathbf{k},\alpha)}=\frac{(1-\alpha)\ket{\psi^I(\mathbf{k})}+\alpha \; \phi(\mathbf{k})^* \ket{\psi^{II}(\mathbf{k})}}{\mathcal{N}},
\label{eq:bloch_one_band}
\end{align}
Here, $\ket{\psi^I(\mathbf{k})}$ and $\ket{\psi^{II}(\mathbf{k})}$ are the Bloch wavefunctions of the valence band for insulators $I$ and $II$ respectively [Eqs.~\eqref{eq:bloch_one_band_1} and~\eqref{eq:bloch_one_band_2}].
$\phi(\mathbf{k})^*=\braket{\psi^{II}(\mathbf{k})|\psi^{I}(\mathbf{k})}$ is the complex conjugate of the overlap between the two Bloch states 
as defined in Eq.~\eqref{eq:overlap_one_band}.
The control parameter $\alpha$ is a real number between $0$ and $1$. The denominator $\mathcal{N}$ is a normalization factor,
\begin{align}
\mathcal{N}=\sqrt{(1-\alpha)^2+\alpha(2-\alpha)|\phi(\mathbf{k})|^2}
\end{align}
which enforces the normalization condition $\braket{\Psi(\mathbf{k},\alpha)| \Psi(\mathbf{k},\alpha)}=1$. It is easy to prove that as long as
the overlap is nonzero $\phi\ne 0$, $\mathcal{N}$ is positive and thus the denominator will not introduce
any singularity. 

When $\alpha=0$, the Bloch state defined above coincides with $\ket{\psi^I(\mathbf{k})}$, 
i.e. the Bloch state for insulator $I$. At $\alpha=1$, the Bloch state becomes that of insulator $II$, up to an unimportant phase factor, 
\begin{align}
&\ket{\Psi(\mathbf{k},\alpha=0)}=\ket{\psi^{I}(\mathbf{k})}
\\
&\ket{\Psi(\mathbf{k},\alpha=1)}=\frac{\phi(\mathbf{k})^*}{|\phi(\mathbf{k})^*|}\ket{\psi^{II}(\mathbf{k})}
\end{align}
Therefore, by varying the parameter $0\le \alpha\le 1$, Eq.~\eqref{eq:bloch_one_band} defines a path between the two insulators.

As proved in Appendix~\ref{sec:symmetry}, 
if insulators $I$ and $II$ preserve certain symmetries (e.g., the time-reversal symmetry, lattice symmetries or
some internal symmetries), the Bloch state $\ket{\Psi(\mathbf{k},\alpha)}$ will preserve the same symmetry. 
In other words, the path defined above preserves all necessary symmetries. 
This is very important for the study of symmetry-protected topological states.

\subsubsection{the insulating gap}
Now, we explore one key problem for the study of adiabatic continuity: {\it is it possible to use the path defined in Eq.~\eqref{eq:bloch_one_band} to deform insulator $I$ into insulator $II$ without closing the insulating gap?}. The answer to this question is yes, 
as long as the wavefunction overlap remains finite for all momenta, $\phi(\mathbf{k})\ne 0$. 
To prove this conclusion, we construct the following hermitian operator,
which will serve as the Hamiltonian for an insulator,
\begin{align}
H(\alpha)=-\sum_{\mathbf{k}}\ket{\Psi(\mathbf{k},\alpha)}\bra{\Psi(\mathbf{k},\alpha)}.
\label{eq:Hamiltonian_one_band}
\end{align}
This Hamiltonian has one control parameter $0\le \alpha\le 1$. It has one flat band with energy $E=-1$
and the Bloch waves for this band is $\ket{\Psi(\mathbf{k},\alpha)}$. All other bands in the system have 
energy $E=0$. If we set the Fermi energy to be between $-1$ and $0$, this Hamiltonian defines a band insulator with one valence band. The band gap for this insulator is $1$.

When $\alpha=0$, the valence band has the same Bloch wavefunctions as insulator $I$, and  for $\alpha=1$, the valence-band Bloch wavefunction coincides with that of insulator $II$. 
For $0<\alpha<1$, the Hamiltonian defines an insulator with a finite insulating gap, and the gap never closes.
As a result, by varying the value of $\alpha$, the Hamiltonian shown in Eq.~\eqref{eq:Hamiltonian_one_band} defines an adiabatic path
between the two insulators.

In the language of topological phase transitions, this observation implies that the two band insulators must belong to the same quantum phase (i.e. have the same topological indices), as long as the wavefunction overlap $\phi(\mathbf{k})$ remains finite for all $\mathbf{k}$.
For two insulators with different topology (i.e. if some topological index takes different values in the two insulators), 
there must be at least one momentum point, at which the overlap vanishes.

\subsubsection{the complex $U(1)$ phase}
\label{sec:U1_for_one_band}
In Eq.~\eqref{eq:bloch_one_band}, we introduced a factor $\phi(\mathbf{k})^*$ in the definition of $\ket{\Psi(\mathbf{k},\alpha)}$. 
This factor is necessary in order to preserve the $U(1)$ phase symmetry, 
which is also known as the $U(1)$ gauge symmetry for band insulators~\cite{Blount1962}. 
In a band insulator, it is known that if we multiply a $U(1)$ phase to a Bloch wavefunction, 
the new wavefunction still describes the same Block state, i.e. $\ket{\psi^I(\mathbf{k})}$ and
 $e^{i\varphi}\ket{\psi^I(\mathbf{k})}$ describe the same Bloch state in insulator $I$, where $\varphi$ is an arbitrary 
$U(1)$ phase. Similarly,  
$\ket{\psi^{II}(\mathbf{k})}$ and $e^{i\varphi'}\ket{\psi^{II}(\mathbf{k})}$
correspond to the same Bloch state in insulator $II$. In other words, when we write down the Bloch states $\ket{\psi^I(\mathbf{k})}$  and $\ket{\psi^{II}(\mathbf{k})}$ for the insulators, there is a freedom to choose an arbitrary phase factor for each of these states. In order to ensure that physical observables [e.g. the Hamiltonian $H(\alpha)$] {\it does not} depend on this arbitrary phase choice, the factor $\phi(\mathbf{k})^*$ is necessary. 
It is straightforward to verify that with the help of this factor,  the Hamiltonian $H(\alpha)$ defined in Eq.~\eqref{eq:Hamiltonian_one_band} is independent of the phase choice, i.e. it is invariant under the transformation
\begin{align}
&\ket{\psi^I(\mathbf{k})}\rightarrow e^{i \varphi} \ket{\psi^I(\mathbf{k})}
\\
&\ket{\psi^{II}(\mathbf{k})}\rightarrow e^{i \varphi'} \ket{\psi^{II}(\mathbf{k})}
\end{align}

In addition, as shown in Appendix~\ref{sec:symmetry}, this factor $\phi(\mathbf{k})^*$ also helps 
to ensure that the adiabatic path
preserves the same symmetries as insulators $I$ and $II$.

\subsection{Insulators with multiple occupied bands}
\label{sec:multi_bands}
Now we consider band insulators with more than one valence bands.  

\subsubsection{wavefunction overlap}
For an insulator with $N$ valence bands, in the non-interacting limit, the ground state wavefunction is
\begin{align}
\ket{\textrm{G}_I} =\prod_{n=1}^N \prod_{\mathbf{k}} c^\dagger_{n, \mathbf{k}} \ket{0}
\label{eq:wavefunction_multiple_bands_1}
\end{align}
Here, we follow the same convention as utilized in Eqs.~\eqref{eq:G_I} and~\eqref{eq:G_II}, except that the creation operators $c^\dagger_{n, \mathbf{k}}$ now have one extra subindex $n$, which labels the valence bands ($n=1,2,\ldots, N$), and $\prod_{n=1}^N$ represents the product for all occupied bands.

Consider another insulator with the same number of valence band, whose ground state wavefunction is 
\begin{align}
\ket{\textrm{G}_{II}} =\prod_{n=1}^N \prod_{\mathbf{k}} d^\dagger_{n, \mathbf{k}} \ket{0}
\label{eq:wavefunction_multiple_bands_2}
\end{align}
where $d^\dagger_{n, \mathbf{k}}$ is the creation operator for the Bloch waves in this insulator.
The quantum overlap between the two ground states of these two insulators factorizes (similar to the case with one valence band)
\begin{align}
|\braket{\textrm{G}_I|\textrm{G}_{II}}|=\prod_{\mathbf{k}} |\phi(\mathbf{k})|
\end{align}
where the Bloch-wave overlap at each momentum point is
\begin{align}
\phi(\mathbf{k})=\braket{0|\prod_{n=1}^N  c_{n, \mathbf{k}}  \prod_{m=1}^N d^\dagger_{m, \mathbf{k}} |0}
\label{eq:overlap_muti_band}
\end{align}
In the first-quantization language, $\phi(\mathbf{k})$ is the determinant of the overlap matrix $\mathcal{F}(\mathbf{k})$
\begin{align}
\phi(\mathbf{k})=\det \mathcal{F}(\mathbf{k})
\label{eq:overlap_muti_band_F_matrix}
\end{align}
where $\mathcal{F}(\mathbf{k})$ is a $N\times N$ matrix with matrix elements
\begin{align}
\mathcal{F}_{n,m}(\mathbf{k})=\braket{0|c_{n, \mathbf{k}} d^\dagger_{m, \mathbf{k}} |0}=\braket{\psi_n^{I}(\mathbf{k})|\psi_m^{II}(\mathbf{k})}
\label{eq:overlap_F_matrix}
\end{align}
where 
\begin{align}
\ket{\psi_n^{I}(\mathbf{k})}=c^\dagger_{n, \mathbf{k}}\ket{0}
\\
\ket{\psi_m^{II}(\mathbf{k})}=d^\dagger_{m, \mathbf{k}}\ket{0}
\end{align}
are the Bloch wavefunctions of the valence bands for insulators $I$ and $II$ respectively, and the subindices $n$ and $m$ are band indices for
valence bands in these two insulators.

We emphasize that the overlap matrix $\mathcal{F}(\mathbf{k})$ is a function of the crystal momentum $\mathbf{k}$. 
However, to simplify the formulas, in this article we will use $\mathcal{F}$ to represent the matrix without showing 
explicitly that this matrix is a function of $\mathbf{k}$.

\subsubsection{the adiabatic path}
\label{sec:adiabatic_path_multiple_bands}

In this section, we will assume that the overlap between the two insulators, i.e. $\phi(\mathbf{k})$ defined in 
Eq.~\eqref{eq:overlap_muti_band}, 
is finite for all momentum points, and then define an adiabatic path between the two insulators. 

According to Eq.~\eqref{eq:overlap_muti_band_F_matrix}, $\phi(\mathbf{k})\ne 0$ implies that the overlap matrix  $\mathcal{F}$ [Eq.~\eqref{eq:overlap_F_matrix}] has a nonzero determinant. As shown in Appendix~\ref{sec:matrices}, 
because $\mathcal{F} \mathcal{F}^\dagger$ is a hermitian matrix, we can find a unitary matrix $\mathcal{U}$, which
diagonalizes $\mathcal{F} \mathcal{F}^\dagger$, 
i.e.  $\mathcal{U}\mathcal{F} \mathcal{F}^\dagger\mathcal{U}^\dagger$ is a diagonal matrix. 
Utilizing the matrices $\mathcal{F}$ and $\mathcal{U}$, 
 we can define $N$ quantum states
\begin{align}
\ket{\Psi_{l}(\mathbf{k},\alpha)}=
\frac{(1-\alpha) \; \mathcal{U}^{*}_{l,n}\ket{\psi_n^I(\mathbf{k})}+\alpha \; \mathcal{U}^{*}_{l,n} \mathcal{F}^*_{n m} \ket{\psi_m^{II}(\mathbf{k})}}{\mathcal{N}_l},
\label{eq:bloch_multi_band}
\end{align}
where $*$ represents complex conjugate; $0\le \alpha \le 1$ is a control parameter and the subindex  $l=1,2,\ldots, N$. 
In this article, we adopt the Einstein summation convention. Unless claimed otherwise, 
repeated band indices  will be summed over, and this sum only goes over all valence bands with band indices between $1$ and $N$, 
while conduction bands (with band indices larger than $N$) will not be included in the sum. 
The denominator $\mathcal{N}_l$ is the normalization factor, which ensures that the quantum state is 
properly normalized, $\braket{\Psi_l|\Psi_l}=1$,  and the value of this normalization function is shown in Eq~\eqref{eq:normalization}. 
In Appendix~\ref{sec:singular_free}, we proved that this normalization factor $\mathcal{N}_l$ never reaches zero, as long as
the overlap is nonzero $\phi(\mathbf{k})\ne 0$, which ensures that  
Eq.~\eqref{eq:bloch_multi_band} is singularity free.

We will prove in the next section that as long as the overlap $\phi(\mathbf{k})$ remains finite,
 the states defined in Eq.~\eqref{eq:bloch_multi_band} are orthonormal 
\begin{align}
\braket{\Psi_{l}(\mathbf{k},\alpha)|\Psi_{l'}(\mathbf{k},\alpha)}=\delta_{l,l'}
\end{align}
As a result, we can design an insulator with $N$ valence bands and utilize these orthonormal states
as the Bloch states of the valence bands, and this insulator will serve as an adiabatic path between insulators $I$ and $II$.
Here, we define the Hamiltonian of this insulator
\begin{align}
H(\alpha)=-\sum_{l=1}^{N} \sum_{\mathbf{k}}  \ket{\Psi_l(\mathbf{k},\alpha)}\bra{\Psi_l(\mathbf{k},\alpha)}
\label{eq:Hamiltonian_multiple_bands}
\end{align}
Because $\ket{\Psi_l(\mathbf{k},\alpha)}$ are orthonormal for $l=1,2,\ldots,N$, 
it is straightforward to verify that $\ket{\Psi_l(\mathbf{k},\alpha)}$ are eigenstates of the Hamiltonian with eigenenergy $E=-1$,
and all other single-particle states orthogonal to $\ket{\Psi_l(\mathbf{k},\alpha)}$ have eigenenergy $E=0$, 
i.e., this Hamiltonian has $N$ (flat) energy bands with energy $E=-1$ and all other energy bands have energy $E=0$.
If the Fermi energy is between $-1$ and $0$, this Hamiltonian defines a band insulator with band gap $\Delta=1$, and 
$\ket{\Psi_l(\mathbf{k},\alpha)}$ are the Bloch waves of the valence bands.
As will be shown in the next section,  for $\alpha=0$ ($\alpha=1$), the ground state wavefunction of this insulator coincides 
with that of insulator $I$ (insulator $II$). And thus $H(\alpha)$ defines an adiabatic path between the two insulators.

\subsubsection{proof for the adiabatic path}
In this section, we prove the conclusions presented in Sec.~\ref{sec:adiabatic_path_multiple_bands}.  
We will first prove that the quantum states defined in Eq.~\eqref{eq:bloch_multi_band} are indeed orthonormal, i.e.,
$\braket{\Psi_{l}(\mathbf{k},\alpha)|\Psi_{l'}(\mathbf{k},\alpha)}
 =\delta_{l,l'}$
Then, we will show that the Hamiltonian defined in Eq.~\eqref{eq:Hamiltonian_multiple_bands} is the Hamiltonian 
for an insulator with $N$ valence bands, and we will further prove that for 
$\alpha=0$  ($\alpha=1$), the ground state recovers that of the insulator $I$ ($II$).

It turns out that it is easier to present the proof using second quantization,
so here we will reformulate the same Bloch states and the Hamiltonian utilizing creation/annihilation operators
defined in Eqs.~\eqref{eq:wavefunction_multiple_bands_1} and~\eqref{eq:wavefunction_multiple_bands_2},
i.e. the creation operator $c^\dagger_{n, \mathbf{k}}$ ($d^\dagger_{m, \mathbf{k}}$) adds one electron
to the $n$th ($m$th) valence band of insulator $I$ ($II$) at crystal momentum $\mathbf{k}$.
Since electrons are fermions, the creation/annihilation operators satisfy the canonical anti-commutation relation
\begin{align}
\{c_{n, \mathbf{k}},c^\dagger_{n', \mathbf{k}'}\}=\delta_{n,n'}\delta_{\mathbf{k},\mathbf{k}'}
\\
\{d_{m, \mathbf{k}}, d^\dagger_{m', \mathbf{k}'}\}=\delta_{m,m'}\delta_{\mathbf{k},\mathbf{k}'}
\end{align}
where $\delta$ is the Kronecker delta. For the anti-commutators between $c$s and $d$s, it is straightforward to prove that 
\begin{align}
\{c_{n, \mathbf{k}},d^\dagger_{m, \mathbf{k}'}\}=\mathcal{F}_{n,m}\delta_{\mathbf{k},\mathbf{k}'}
\label{eq:commutator_c_and_d_1}
\\
\{d_{m, \mathbf{k}}, c^\dagger_{n, \mathbf{k}'}\}=\mathcal{F}^*_{n,m}\delta_{\mathbf{k},\mathbf{k}'}
\label{eq:commutator_c_and_d_2}
\end{align}
(See Appendix~\ref{sec:commutator_c_and_d} for details).

Utilizing these creation and annihilation operators, as well as the matrices $\mathcal{F}$ and $\mathcal{U}$ defined in 
Sec~\ref{sec:adiabatic_path_multiple_bands}, we can define creation operators
\begin{align}
a^\dagger_{l,\mathbf{k}}=
\frac{(1-\alpha) \; \mathcal{U}^{*}_{l,n}c^\dagger_{n,\mathbf{k}}+\alpha \; \mathcal{U}^{*}_{l,n} \mathcal{F}^*_{n m} d^\dagger_{m,\mathbf{k}}}{\mathcal{N}_l},
\label{eq:a_operator}
\end{align}
Here, repeated indices are summed over, same as in Eq.~\eqref{eq:bloch_multi_band}.
It is straightforward to verify that this creation operator creates the Bloch state $\ket{\Psi_{l}(\mathbf{k},\alpha)}$ defined 
in Eq.~\eqref{eq:bloch_multi_band}, i.e. $\ket{\Psi_{l}(\mathbf{k},\alpha)}=a^\dagger_{l,\mathbf{k}}\ket{0}$.
In Appendix~\ref{sec:commutators_a}, we proved that as long as the overlap $\phi(\mathbf{k})$ is nonzero,
these $a^\dagger_{l,\mathbf{k}}$ operators, and the corresponding
annihilation operators, satisfies canonical anti-commutation relations
\begin{align}
\{a_{l,\mathbf{k}},a^\dagger_{l',\mathbf{k}}\}=\delta_{l.l'}
\label{eq:commutator_for_as}
\end{align}
The anti-commutation relation implies that the quantum states defined in Eq.~\eqref{eq:bloch_multi_band} are orthonormal, 
because
\begin{align}
\delta_{l,l'}=\braket{0|\{a_{l,\mathbf{k}},a^\dagger_{l',\mathbf{k}}\}|0}
=\braket{\Psi_{l}(\mathbf{k},\alpha)|\Psi_{l'}(\mathbf{k},\alpha)}
\end{align}

Now we examine the Hamiltonian defined in Eq.\eqref{eq:Hamiltonian_multiple_bands} and rewrite it in the second-quantization language
\begin{align}
H(\alpha)=-\sum_{l=1}^N\sum_{\mathbf{k}} a^\dagger_{l,\mathbf{k}} a_{l,\mathbf{k}}
\end{align}
Along with the anti-commutation relation [Eq.~\eqref{eq:commutator_for_as}], it is easy to verify that this Hamiltonian describes a band insulator 
with $N$ valence bands. $a^\dagger_{l,\mathbf{k}}$ are the creation operators for the Bloch states in the valence bands ($l=1,2,\ldots, N$). 
All the valence bands in this insulator have energy $-1$, while the conduction bands have energy $0$. Here, we set the Fermi energy 
into the band gap, i.e., between $-1$ and $0$. For any values of $0\le \alpha\le 1$, the insulating gap never closes and the value remains $1$.

For  $\alpha=0$, we know from Eq.~\eqref{eq:a_operator} that
\begin{align}
a^\dagger_{l,\mathbf{k}}=\mathcal{U}^{*}_{l,n} c^\dagger_{n,\mathbf{k}}
\end{align}
Because $\mathcal{U}$ is a unitary matrix (i.e. $\mathcal{U}^{*}_{l,n} \mathcal{U}_{l,n'}=\delta_{n,n'}$), the Hamiltonian at $\alpha=0$
is
\begin{align}
H(\alpha=0)=-\sum_{n=1}^N\sum_{\mathbf{k}} c^\dagger_{n,\mathbf{k}} c_{n,\mathbf{k}}
\end{align}
Therefore, the ground state is identical to that of insulator $I$, i.e.,
all Bloch states created by $c^\dagger_{n,\mathbf{k}}$ for $n=1,2,\ldots, N$
are occupied.

For $\alpha=1$, Eq.~\eqref{eq:a_operator} implies that
\begin{align}
a^\dagger_{l,\mathbf{k}}=\frac{1}{\mathcal{N}_l}\mathcal{U}^{*}_{l,n}\mathcal{F}^{*}_{n,m}d^\dagger_{m,\mathbf{k}}.
\end{align}
Thus the Hamiltonian becomes
\begin{align}
H(\alpha=1)=-\sum_{\mathbf{k}} 
\frac{\mathcal{F}^{*}_{n,m}\mathcal{U}^{*}_{l,n}\mathcal{U}_{l,n'}\mathcal{F}_{n',m'}
}{\mathcal{N}_l^2}d^\dagger_{m,\mathbf{k}} d_{m',\mathbf{k}}
\label{eq:Hamiltonian_alpha_1_partial_simplified}
\end{align}
As proved in Appendix~\ref{sec:matrices},
\begin{align}
\frac{\mathcal{F}^{*}_{n,m}\mathcal{U}^{*}_{l,n}\mathcal{U}_{l,n'}\mathcal{F}_{n',m'}}{\mathcal{N}_l^2}=\delta_{m,m'}
\end{align}
and thus this Hamiltonian can be simplified 
\begin{align}
H(\alpha=1)=-\sum_{m=1}^N\sum_{\mathbf{k}}d^\dagger_{m,\mathbf{k}} d_{m,\mathbf{k}}.
\end{align}
The ground state for this Hamiltonian coinides with that of the insulator $II$, i.e.,
all Bloch states created by $d^\dagger_{m,\mathbf{k}}$ for $m=1,2,\ldots, N$
are occupied.

\subsubsection{insulators with different numbers of valence bands}
Consider two insulators with different numbers of valence bands. It is easy to realize that these two insulators are 
{\it not} adiabatically connected, because it is impossible to change the number of valence bands in 
a band insulator without going through a gapless (metallic) state. 

At the same time, we know that the overlap function also vanishes. Utilizing the overlap function 
defined in Eq.~\eqref{eq:overlap_muti_band}, we know that
\begin{align}
\phi(\mathbf{k})=\braket{0|\prod_{n=1}^N  c_{n, \mathbf{k}} \prod_{m=1}^{N'} d^\dagger_{m, \mathbf{k}} |0}
\end{align}
where $N$ and $N'$ are the number of valence bands for the two insulators respectively. It is transparent that 
$\phi(\mathbf{k})=0$, if $N \ne N'$.

In summary, for two insulators with different numbers of valence bands, the two insulators are not adiabatically
connected, and the wavefunction overlap is zero.

\subsection{Symmetry protected topological states}
As mentioned above and proved in Appendix~\ref{sec:symmetry}, if insulators $I$
 and $II$ preserves certain symmetry, the adiabatic path that we defined will preserve the same symmetry.
This property is very important for the study of symmetry-protected topological states,
where the topological index can only be defined in the presence of certain symmetries.
There, when we discuss adiabatic paths that connect two quantum states, 
we must ensure that the symmetry that are utilized to 
define the topological index is preserved along the path. And 
the adiabatic path that we constructed above indeed preserves the symmetry, as long as the symmetry 
is preserved in insulators $I$ and $II$.

\subsection{Insulators with different lattice structures}
In the previous sections, we assumed that the two insulators ($I$ and $II$) have the same Brillouin zone, and thus
we can use the same momentum points in both insulators to compute the wave-function overlap. This assumption is not necessary, and all the conclusions above
can be generalized, even if two insulators have different lattice structures, and thus different Brillouin zones.

This is because the topology of a band insulator remains invariant as we adiabatically deform the lattice structure,
as long as the gap remains finite 
(For certain topological states, e.g. topological crystalline insulators~\cite{Fu2011},  the symmetry of the underlying 
lattice plays an essential role in the definition of the topological structure. There, as long as
the deformation of the lattice structure preserves the essential symmetry, the topological structure also remains invariant).
Thus, we can deform adiabatically the crystal structure of one insulator into the structure of the other insulator, 
and then all the conclusions above can be generalized.

Finally, we emphasize that the adiabatic deformation discussed here is not unique. Instead, there exists infinite many different paths to deform the crystal structure. As long as the deformation is adiabatic, our conclusion will remain the same. 

Below, in Sec.~\ref{sub:section}, we will provide
one example on how to compare the Bloch waves in two insulators with different lattice structures.

\subsection{Adiabatic band flattening}
Above, we defined a Hamiltonian with flat bands to demonstrate the adiabatic continuity. This band structure (with flat bands)
are different from that of a real insulator, where the energy bands are in general not flat and not degenerate.
However, for the study of adiabatic continuity and/or topological phase transitions, this difference doesn't play any essential role.
This is because in an arbitrary band insulator, we can adiabatically flatten all the bands and adjust the energy of each band without
changing the Bloch wavefunctions. The adiabatic flattening of energy bands are widely utilized in the study of topological 
insulator/superconductors, and it is known that topological properties remain invariant as we flatten the bands in a band insulator, 
as long as the band gap remains open (See for example Refs.~\onlinecite{kitaev2009art} and \onlinecite{schnyder2008}).

\subsection{Examples}
\label{sub:section}
\begin{figure}
 \includegraphics[width=0.85\linewidth]{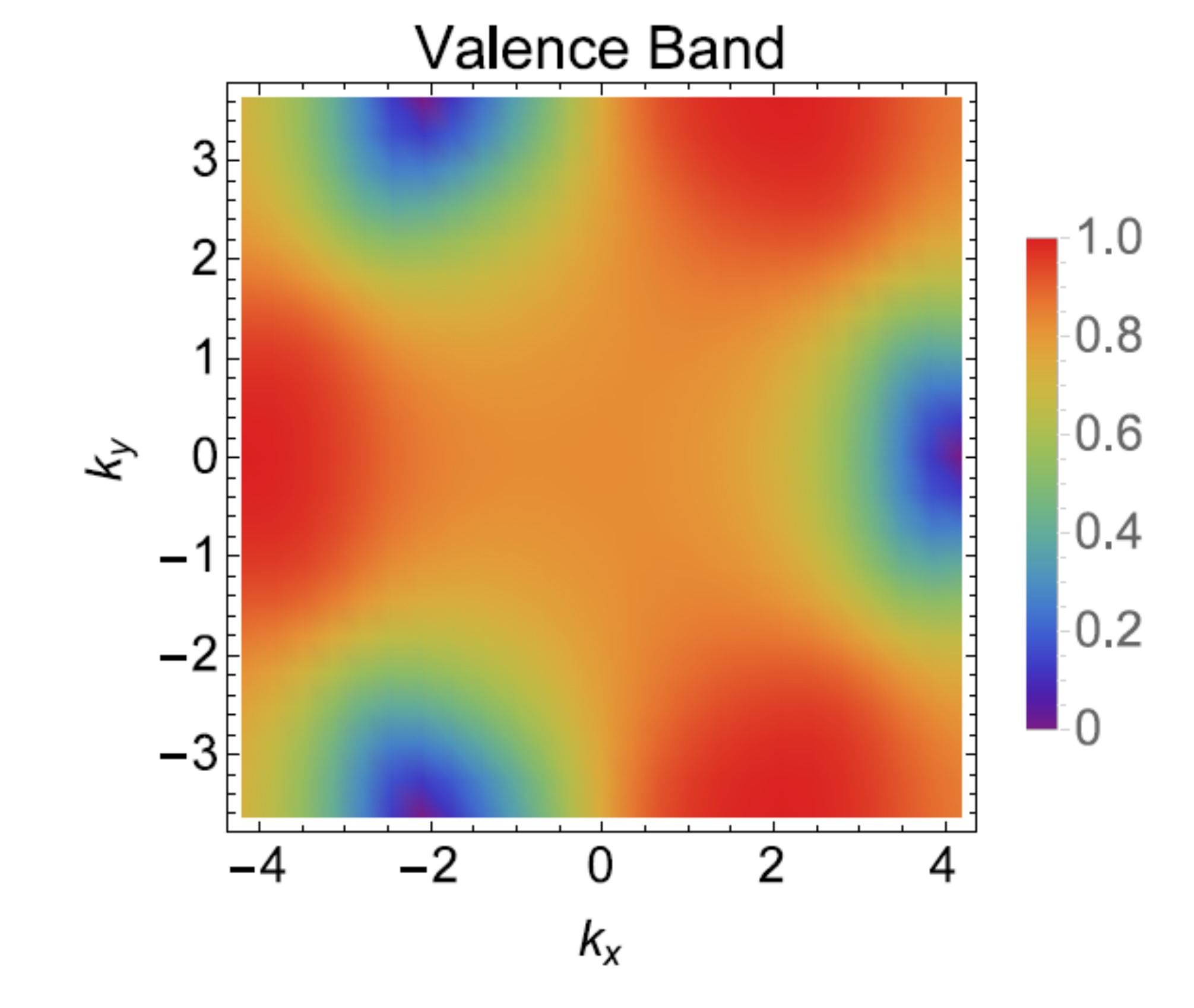}
\caption{The absolute value of the Bloch-wavefunction overlap in Haldane's model. Here, we examined two insulating states
in the model of Haldane with different Chern numbers ($+1$ and $0$). Utilizing the Bloch waves of the valence bands
in the two insulators, we computed the wavefunction overlap $\phi(\mathbf{k})$ and plotted its absolute value as a function of
the crystal momentum $k_x$ and $k_y$. As shown in the figure, the overlap vanishes at certain momentum point, which happens
to be the $K$ point for this model.}
\label{fig:Haldane}
\end{figure}

In this section, we present examples to demonstrate that for two insulators with different topology,  
the Bloch-wavefunction overlap must vanish at certain momentum point in the Brillouin zone.

\subsubsection{Insulators with different topology}
First, we consider insulators with different topological structures and show that the wavefunction overlap must vanish at some moment point. We start by considering the model of Haldane~\cite{haldane1988}. As pointed out by Haldane, 
for a honeycomb lattice, the Dirac band-touching point can be gapped by two different methods: (1) introducing
a magnetic flux pattern, which breaks the time-reversal symmetry or (2) introducing a staggered potential, which
breaks the degeneracy between the two sublattices. At half-filling, these two approaches result in  two different insulators
with different topology, a topologically-nontrivial Chern insulator and a topologically-trivial conventional insulator.

Utilizing these two topologically different insulators, we can compute the overlap between Bloch states in their valence bands,
i.e., $\phi(\mathbf{k})$ defined above.
As shown in Fig.~\ref{fig:Haldane}, this overlap vanishes at the $K$ point, in agreement with our conclusions above. 

For Chern insulators with different Chern numbers, zero wavefunction overlap has been observed and proved in earlier studies using other approaches~\cite{Yang2013, Huang2016}. Our theorem indicates that the same conclusions will remain for any types of topological indices, including symmetry-protected topological states. To demonstrate this conclusion, we have also computed the wavefunction overlap in other models with one or more valence bands (not shown), e.g. the Kane-Mele model~\cite{kane2005} and the Bernevig-Hughes-Zhang model~\cite{bernevig2006}. For insulating states with different topology,  we always find some momentum point, at which the wavefunction overlap $\phi(\mathbf{k})$ reaches zero.

\subsubsection{Topologically equivalent insulators with different lattice structures}
\begin{figure}
\includegraphics[width=0.85\linewidth]{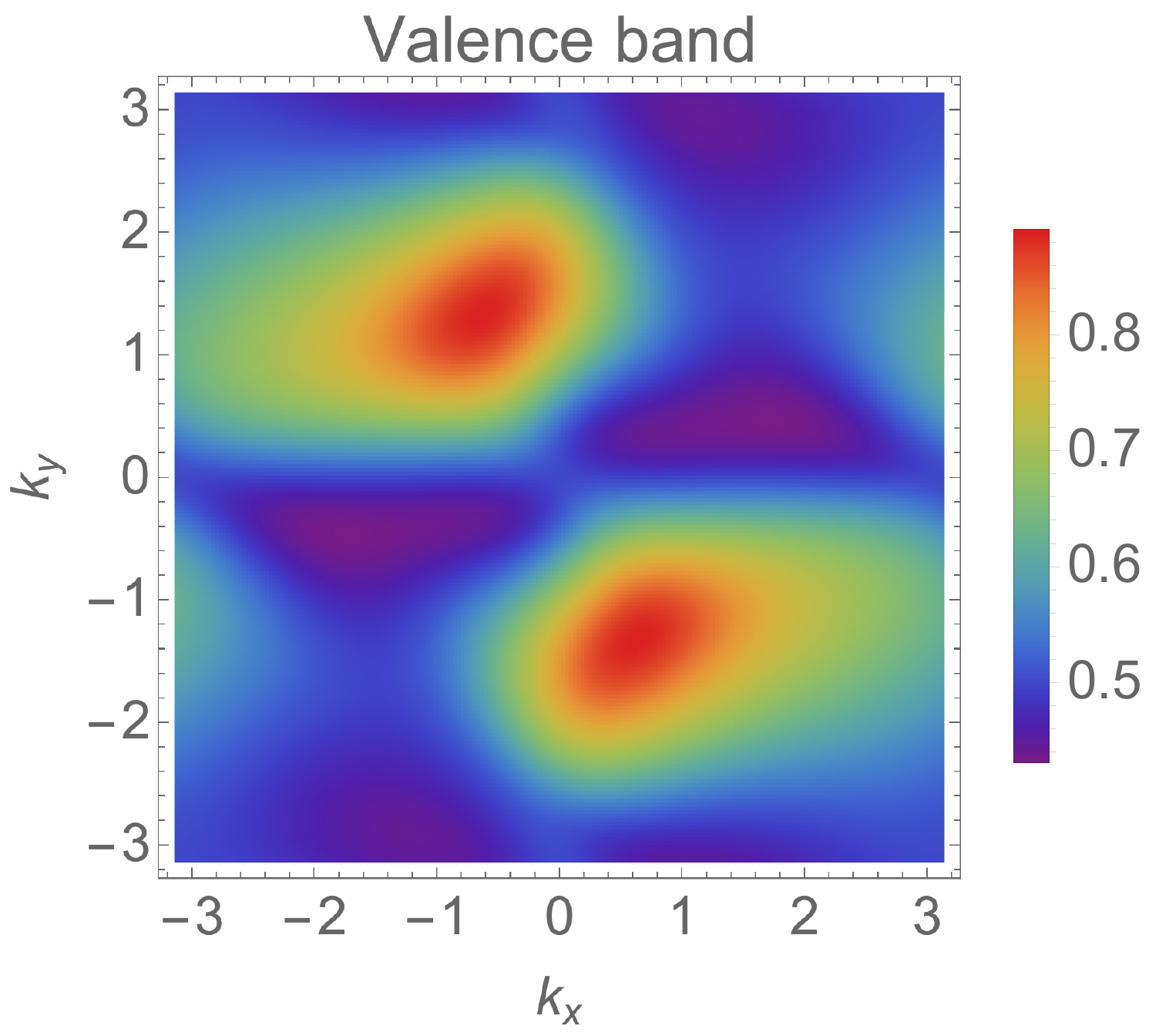}
\caption{The absolute value of the Bloch-wavefunction overlap between 
the Kane-Mele model and the Bernevig-Hughes-Zhang model.
Here, we compute the wavefunction overlap for the quantum spin Hall insulators 
described by the Kane-Mele model and the Bernevig-Hughes-Zhang model.
Because the two models have different Brillouin zone, here we used a continuous 
mapping to map the Brillouin zone of the Kane-Mele model to that of the Bernevig-Hughes-Zhang model. The plot shows the absolute value of the overlap as a function of 
the crystal momentum $(k_x,k_y)$. As shown in the figure, the overlap remains finite indicating that
these two insulators are topologically equivalent.}
\label{fig:KMandBHZ}
\end{figure}

Here, we consider two topologically equivalent insulators with different lattice structures. 
In this example, we compare the quantum spin Hall insulators in the Kane-Mele model~\cite{kane2005} and the Bernevig-Hughes-Zhang model~\cite{bernevig2006}.

These two models assum very different lattice structures (honeycomb and square) and thus
the Brillouin zones of these two models have very different geometry. As shown in Appendix (Sec.~\ref{sec:differentmodel}), 
we can use a continuous one-to-one correspondence 
to map the Brillouin zone of the Kane-Mele model 
to that of the Bernevig-Hughes-Zhang model. (There exist infinite many such mappings, and here we just adopt one of them to demonstrate the physics). As shown in Fig.~\ref{fig:KMandBHZ}, despite differences in lattice structures etc., the quantum-spin-Hall insulators described by 
these two different models show finite wavefunction overlap, which implies immediately that they are topologically equivalent.

\section{interacting systems}
\label{sec:interacting}
In the presence of interactions, we can no longer utilize (decoupled) single-particle (Bloch) states to characterize the 
ground state of a many-body quantum system. However, we can prove a similar theorem for generic quantum systems, 
which reveals a universal relation between adiabatic continuity and the wavefunction overlap.
\begin{theorem}
For any two quantum states with nonzero overlap, i.e., $\ket{\psi}$ and $\ket{\psi'}$ with $\braket{\psi|\psi'}\ne 0$,  
a Hamiltonian $H(\alpha)$ can be defined, such that by tuning the control parameter $\alpha$, the ground state 
of the Hamiltonian evolves adiabatically from $\ket{\psi}$ to $\ket{\psi'}$. During this adiabatic procedure, the energy 
gap between the ground and excited states remains finite.
\end{theorem}

It must be emphasized that although this theorem shares some similarities with what was discussed above
for band insulators (and the proof is along the same line of thinking as will be shown below), 
this theorem is fundamentally different from the conclusions shown in the
previous section. This theorem covers a wider range of systems (interacting and non-interacting),
but it is a weaker statement in comparison to what we have proved in the previous section for band insulators.
For non-interacting band insulators, we showed that the adiabatic path can be achieved using
a {\it non-interacting Hamiltonian}. But for more general situations considered in the theorem above,
the Hamiltonian that describes the adiabatic path may contain interactions, i.e., 
we have to enlarge the scope of Hamiltonians in order to construct the adiabatic path for generic systems.
Proving that two states are connected by a {\it non-interacting} Hamiltonian is 
a stronger statement than proving that they are connected by a Hamiltonian, 
without the non-interacting constraint.  Another way to see this difference is by examining 
the adiabatic path. As will be shown below, the Hamiltonian that we constructed to prove this theorem 
contains interactions. Even in the non-interacting limit, in general, it will not recover the 
non-interacting Hamiltonian 
utilized in the previous section.

In this section, we prove this theorem, and its implications for quantum phase transitions will be 
discussed in the next section. As will be shown in the next section, for topological phase transitions, there exist major differences between interacting and non-interacting systems.
In particular, in the presence of strong interactions, the connection between our theorem and 
quantum phase transitions becomes much more complicated in comparison to non-interacting 
systems discussed in the previous section. As a result, we can only apply this theorem for the 
study of certain interacting topological systems. 

\subsection{Adiabatic path connecting two quantum states}
Consider two quantum states $\ket{\psi}$ and $\ket{\psi'}$. Here $\ket{\psi}$ and $\ket{\psi'}$ are 
generic quantum states, instead of single-particle states. We can define overlap between the two states as
\begin{align}
\phi=\braket{\psi|\psi'}
\end{align}
Define a new quantum state
\begin{align}
\ket{\Psi(\alpha)}=\frac{(1-\alpha)\; \ket{\psi}+\alpha\; \phi^* \ket{\psi'}}{\mathcal{N}},
\label{eq:wavefunction}
\end{align}
where $0\le\alpha \le 1$ is a real number between $0$ and $1$ and $\phi^*$ is the complex conjugate of the wavefunction overlap.
The denominator $\mathcal{N}$ is a normalization factor,
\begin{align}
\mathcal{N}=\sqrt{(1-\alpha)^2+\alpha(2-\alpha)|\phi|^2}
\end{align}
which ensures the normalization condition $\braket{\Psi(\alpha)| \Psi(\alpha)}=1$. 
Utilizing this wavefunction, we can define a hermitian quantum operator
\begin{align}
H(\alpha)=-\ket{\Psi(\alpha)}\bra{\Psi(\alpha)},
\label{eq:Hamiltonian_interacting}
\end{align}
and this quantum operator will serve as our Hamiltonian.

If $H(\alpha)$ is a Hamiltonian and $\alpha$ is a control parameter, the energy spectrum of the system can be figured out immediately. The ground state of the system is $\ket{\Psi(\alpha)}$ with eigenenergy $-1$
\begin{align}
H(\alpha)\ket{\Psi(\alpha)}=-\ket{\Psi(\alpha)}\braket{\Psi(\alpha)|\Psi(\alpha)}=-\ket{\Psi(\alpha)},
\label{eq:Hamiltonian}
\end{align}
All other eigenstates of $H$ have eigenenergy $0$, which are the excited states. In other words, this Hamiltonian defines a gapped system with a unique ground state, while all the excited states are separated by an energy gap.

When $\alpha=0$, the ground state is $\ket{\Psi(0)}=\ket{\psi}$. At $\alpha=1$, the ground state is $\ket{\Psi(1)}=\ket{\psi'}$ up to a phase factor. 
For $0<\alpha<1$, the energy gap between the ground and excited states always remain finite ($\Delta=1$), and thus as we tune $\alpha$ from
$0$ to $1$, it offers an adiabatic path to deform (adiabatically) a quantum state $\ket{\psi}$ into a different quantum state $\ket{\psi'}$ 
without closing the excitation gap. 

For quantum phase transitions, the existence of such an adiabatic path implies that $\ket{\psi}$ and $\ket{\psi'}$ belongs to the same quantum phase, i.e. we 
can go from one to the other without going through a quantum phase transition. This conclusion remains valid as long as the overlap remains finite $\braket{\psi|\psi'}\ne 0$.

As shown in Appendix~\ref{sec:symmetry}, this adiabatic path preserves the same symmetry as $\ket{\psi}$ and $\ket{\psi'}$.

\subsection{$U(1)$ phase symmetry}
In Eq.~\eqref{eq:wavefunction}, a factor $\phi^*=\braket{\psi'|\psi}$ is introduced in the definition of $\ket{\Psi(\alpha)}$. 
This factor is necessary in order to preserve the $U(1)$ phase symmetry.
Because the proof is in strong analogy to the non-interacting case discussed
discussed in Sec.~\ref{sec:U1_for_one_band}, here we will not repeat the analysis,
and it is straightforward to verify that with this $\braket{\psi'|\psi}$ factor,  
$H(\alpha)$ is invariant under the transformation
\begin{align}
&\ket{\psi}\rightarrow e^{i \phi} \ket{\psi}
\\
&\ket{\psi'}\rightarrow e^{i \phi'} \ket{\psi'}
\end{align}
In addition, as shown in Appendix~\ref{sec:symmetry}, this factor $\phi^*$ also helps 
to ensure that the adiabatic path preserves the same symmetries as $\ket{\psi}$ and $\ket{\psi'}$.

\section{applications to quantum phase transitions}
\label{sec:interaction_phase_transition}
For the study of quantum phase transitions, this theorem has two immediate implications: (1) if two quantum states belong to two different quantum 
phases, and it is impossible to go from one to the other adiabatically without going through a quantum phase transition point, 
the overlap between the two quantum wavefunctions must be strictly zero, i.e. the two wavefunction must be orthogonal to each other; and (2) if two 
quantum states have finite overlap, they must belong to the same quantum phase, i.e., one can turn a state into the other adiabatically without going 
through a quantum phase transition. 

This observation enforces a strong constraint on quantum wavefunctions in different quantum phases. However, before we can apply this knowledge to the study of 
quantum phase transitions, one challenge has to be resolved, the {\it orthogonality catastrophe}.
Based on the orthogonality theorem from Anderson, in the thermodynamic limit, the overlap between two different quantum wavefunctions 
shall vanish due to the infinite degrees of freedom~\cite{Anderson1967}.  
To utilize the theorem discussed above to study quantum phase transitions, it is necessary to find a way to distinguish 
zero overlap caused by Anderson's orthogonality theorem and zero overlap caused by the absence of an adiabatic path.
There are three ways to take care of the orthogonality catastrophe:
\begin{itemize}
\item Utilizing another zero to cancel the zero induced by the orthogonality theorem. One technique that can achieve this objective
is the strange correlator as shown in Ref.~\onlinecite{You2014}.
\item Separate an infinite system into smaller subsystems with finite degrees of freedom, and then investigate the overlap in each subsystem, 
which doesn't suffer from the orthogonality catastrophe. This technique is applicable for non-interacting systems and certain interacting systems.
\item Study finite-size systems and then extrapolate to the infinite-size limit via finite-size scaling. 
This last approach is directly relevant to numerical studies.
\end{itemize}
Below, we will explore some examples to demonstrate the second and the third techniques.

\subsection{Quantum Hall and Chern insulators}
For certain topological states, the topological structure is well defined for both finite and infinite systems. 
The most well-known example of this type is the integer and fractional quantum Hall systems, as well as the integer
and fractional Chern insulators, where the topological index can be computed using twisted boundary conditions
for both finite-size and infinite systems~\cite{niu1985}.

\subsubsection{definition of topological indices for a finite-size system}
Consider a finite-size two-dimensional many-body systems with size 
$L_x\times L_y$. We enforce twisted boundary conditions for many-body wavefunctions
\begin{align}
&\psi(\ldots,x_i+L_x,y_i,\ldots)
=e^{i \varphi_x}\psi(\ldots, x_i,y_i,\ldots)
\\
&\psi(\ldots,x_i,y_i+L_y,\ldots)=e^{i \varphi_y}\psi(\ldots, x_i,y_i,\ldots)
\end{align}
where $\psi$ is a many-body wavefunction, while $x_i$ and $y_i$ are the $x$ and $y$ cooridnates of the $i$th particle. $\varphi_x$ and $\varphi_y$ are two phase factors. 
For $\varphi_x=\varphi_y=0$ ($\varphi_x=\varphi_y=\pi$), it recovers the periodic (anti-periodic) boundary conditions. For other values of $\varphi_x$ and $\varphi_y$, it is known
as the twisted boundary conditions.

We can find the ground state of a quantum system under twisted boundary conditions $\ket{\psi(\varphi_x,\varphi_y)}$. In general, the ground state wavefunction depends on 
the values of $\varphi_x$ and $\varphi_y$. For a gapped system, we can define the following integral
\begin{align}
  C = \int_0^{2\pi} d\varphi_x \int_0^{2\pi} d\varphi_y 
\frac{ \braket{\partial_{\varphi_x}
      \psi | \partial_{\varphi_y} \psi} - \braket{\partial_{\varphi_y}
      \psi | \partial_{\varphi_x} \psi}}{2 \pi i}
  \label{eq:Chern_number}
\end{align}
As pointed out in Ref.~\onlinecite{niu1985}, this integral is a topological invariant, i.e. the first Chern number, regardless of the size of the system.
In the thermodynamic limit, this topological index coincides with the Hall conductivity~\cite{niu1985}. Because
the definition utilizes many-body wavefunctions (without using single-particle Bloch waves), it is applicable for both interacting 
and non-interacting systems. In the non-interacting limit, it recovers the Chern number computed using single-particle Bloch 
waves~\cite{thouless1982}.

It is also worthwhile to mention that it is straightforward to generalize this definition to fractional quantum Hall systems 
and fractional Chern insulators. Once topological degeneracy is taken into account, the integral shown above produces 
fractional values, i.e. the fractional Hall conductivity~\cite{Sheng2003}.

\subsubsection{wavefunction overlap and topological index}
Consider a 2D finite-size system with Hamiltonian $H_1$ and another 2D system with the same size but a different Hamiltonian $H_2$. 
Here, we allow the Hamiltonians to contain interactions, and we assume 
that the ground states are gapped for both Hamiltonians (for any twisted boundary conditions).
 We can find the many-body ground states for the two Hamiltonians under twisted boundary condition $\ket{\psi_1(\varphi_x,\varphi_y)}$
and  $\ket{\psi_2(\varphi_x,\varphi_y)}$ respectively. Using Eq.~\eqref{eq:Chern_number}, one 
can compute the Chern number for the ground states of both Hamiltonians.

Here, we ask the following question: 
{\it if the ground states of the two Hamiltonians have different Chern numbers, 
what is the wavefunction overlap between the two insulators, $\braket{\psi_1(\varphi_x,\varphi_y)|\psi_2(\varphi_x,\varphi_y)}$}. 
Because we have set the system-size to finite, the wavefunction overlap
{\it does not} suffer from the orthogonality catastrophe, and thus we can directly apply the theorem proved above.

Because the two ground states have different Chern numbers, it is impossible to adiabatically deform one state 
into the other without closing the excitation gap (between the ground state and the first excited state). 
This implies that no matter how we try to deform $H_1$ into $H_2$, adiabatically, the excitation gap must close
for at least one set of $\varphi_x$ and $\varphi_y$. Utilizing the theorem proved above, 
this implies that we can find at least one set of  $\varphi_x$ and $\varphi_y$, the wavefunction overlap 
vanishes $\braket{\psi_1(\varphi_x,\varphi_y)|\psi_2(\varphi_x,\varphi_y)}=0$. 
Otherwise, an adiabatic path will exist, which is in contradiction to the assumption that the two states have different 
Chern numbers.

Now we consider the opposite situation, where $\braket{\psi_1(\varphi_x,\varphi_y)|\psi_2(\varphi_x,\varphi_y)}\ne 0$ for all 
possible values of $\varphi_x$ and $\varphi_y$. Utilizing the theorem shown above, for any twisted boundary condition,
we can construct an adiabatic path between these two quantum states without closing the gap. As a result, 
the two states must have the same Chern number.

\subsubsection{topological phase transitions in interacting systems}
Now we study a topological phase transitions in a 2D interacting system. Consider a Hamiltonian $H(\alpha)$, 
where $\alpha$ is a control parameter. We assume that by tuning the control parameter $\alpha$, the system undergoes a topological phase transition, where the Chern number changes its value, i.e.,  the Hamiltonian has a gapped ground state for both 
$\alpha>\alpha_C$ and $\alpha<\alpha_C$, but the ground states have different Chern numbers for $\alpha>\alpha_C$ and $\alpha<\alpha_C$.
Here again, we consider a finite-size system, although one can take the thermodynamic limit later via finite size scaling. As shown above and 
pointed out in Ref.~\onlinecite{Varney2011}, even for finite size systems, the Chern number and the topological phase transition is well-defined.

The ground-state wavefunction of  this Hamiltonian, $\ket{\psi_\alpha(\varphi_x,\varphi_y)}$ 
depends on the value of the control parameter $\alpha$, as well as the phases of 
the twisted boundary conditions $\varphi_x$ and $\varphi_y$.
We can compute the wavefunction overlap for the ground states at different values of $\alpha$,
\begin{align}
\phi_{\alpha_1,\alpha_2}(\varphi_x,\varphi_y)=\braket{\psi_{\alpha_1}(\varphi_x,\varphi_y)|\psi_{\alpha_2}(\varphi_x,\varphi_y)}
\end{align}
The conclusions that we proved above indicate immediately that if this overlap never vanishes for any $\varphi_x$ and $\varphi_y$, 
$H(\alpha_1)$ and $H(\alpha_2)$ describe states in the same quantum phase, i.e. $\alpha_1>\alpha_C$ and $\alpha_2>\alpha_C$, 
or $\alpha_1<\alpha_C$ and $\alpha_2<\alpha_C$.

Similarly. if we compute the overlap for two wavefunctions from two different topological phases,  (e.g., $\alpha_1>\alpha_C$ and $\alpha_2<\alpha_C$), then this overlap must vanish for some values of $\varphi_x$ and $\varphi_y$.
A special case of this type has been shown in Ref.~\onlinecite{Varney2011}, where $\alpha_1$ and $\alpha_2$ are very close to the transition point, 
i.e. $\alpha_1=\alpha_C+\epsilon$ and $\alpha_2=\alpha_C-\epsilon$ where $\epsilon$ is a very small positive number.
There, the vanishing wavefunction overlap results in a singularity (i.e. a Dirac $\delta$-function) in the fidelity 
matrix~\cite{zanardi2006, campos2007,rigol2009}, 
which can be used to pin-point the topological phase transition in a finite-size interacting system. 
The results shown above generalize the same conclusion for any values of $\alpha_1>\alpha_C$ and $\alpha_2<\alpha_C$, 
close or far away from the topological transition point.

\subsection{Factorized wavefunction overlap in certain interacting systems}
In general, a many-body ground-state wavefunction of an interacting system cannot be factorized as 
the product of single-particle (or few-particle) wavefunctions, 
in contrast to non-interacting systems discussed in Sec.~\ref{sec:band_insulator}.
However, for certain interacting systems, such a factorization could happen, 
which offers us another way to avoid the orthogonality catastrophe in the study of wavefunction overlap.

Here we consider a (AA-stacked) bilayer Kane-Mele model as studied in Ref.~\onlinecite{He2016a}. 
For each layer, we have a non-interacting Kane-Mele model (on a honeycomb lattice), 
which describes a $Z_2$ topological insulator. Between the layers, 
an interlayer anti-ferromagnetic spin-spin interaction is introduced between interlayer nearest neighbors.

In this model, because the $z$-component of the spin is conserved, the insulating ground state is 
characterized by an integer-valued topological index, known as the spin Chern number. 
In the non-interacting limit, the topological index is $+2$, i.e., the system is topologically nontrivial.
Because there is no interaction, the ground state factorizes as the anti-symmetrized product of Bloch states
\begin{align}
\ket{\psi_I}=\prod_\mathbf{k} c_{\textrm{t},\mathbf{k}}^\dagger d^{\dagger}_{t,\mathbf{k}} c^\dagger_{\textrm{b},\mathbf{k}} d^\dagger_{\textrm{b},\mathbf{k}}\ket{0}
\end{align}
where $c_{\textrm{t},\mathbf{k}}^\dagger$ and $d^{\dagger}_{\textrm{t},\mathbf{k}}$ are the 
creation operators for the two valence bands in the top layer. Here, the top layer is a non-interacting Kane-Mele model, 
which has two valence bands (taking into account the spin degrees of freedom).
The other two creation operators $c^\dagger_{\textrm{b},\mathbf{k}}$ and $d^\dagger_{\textrm{b},\mathbf{k}}$ 
are for the bottom layer, which is identical to the top layer.

When the interlayer anti-ferromagnetic coupling is infinitely strong, electrons between the two layers
form singlet pairs (i.e., dimers). At half-filling, the dimers fill up the whole system, and electrons can no 
longer move, i.e. the system becomes a topologically-trivial insulator with spin Chern number $0$.
Here, the ground state wavefunction is
\begin{align}
&\ket{\psi_{II}}=\nonumber\\
&\prod_i (a_{\textrm{t},i,\uparrow}^\dagger a_{\textrm{b},i,\downarrow}^\dagger-a_{\textrm{t},i,\downarrow}^\dagger a_{\textrm{b},i,\uparrow}^\dagger)
(b_{\textrm{t},i,\uparrow}^\dagger b_{\textrm{b},i,\downarrow}^\dagger-b_{\textrm{t},i,\downarrow}^\dagger b_{\textrm{b},i,\uparrow}^\dagger)
\ket{0}
\end{align}
Here, $a^\dagger$ and $b^\dagger$ are the creation operator for the $A$ and $B$ sublattices of the honeycomb lattice respectively. The subindices $\textrm{t}$ and
$\textrm{b}$ represent the top and bottom layers, and $i$ is the index for unit cells. $\uparrow$ and $\downarrow$ are spin indices
 (spin up and down).
Here, $a_{\textrm{t},i,\uparrow}^\dagger a_{\textrm{b},i,\downarrow}^\dagger-a_{\textrm{t},i,\downarrow}^\dagger a_{\textrm{b},i,\uparrow}^\dagger$
 and $b_{\textrm{t},i,\uparrow}^\dagger b_{\textrm{b},i,\downarrow}^\dagger-b_{\textrm{t},i,\downarrow}^\dagger b_{\textrm{b},i,\uparrow}^\dagger$
 create spin singlets (dimers) in the $A$ and $B$ sites of the $i$th unit cell.

Because the non-interacting limit and  the strong-coupling limit have different topological indices ($+2$ and $0$),
a topological phase transition must arise as the  anti-ferromagnetic coupling strength increases.
This transition was observed and studied using quantum Monte Carlo simulations~\cite{He2016a}.

Here, we focus on the non-interacting limit and the infinite-coupling limit. As shown above,
in both cases, the ground states are product states. With periodic boundary conditions, 
the number of momentum points in a Brillouin zone coincides with the number of unit cells in the real space. 
Thus, a one-to-one correspondence can be defined between the unit cell index $i$ and crystal momentum $\mathbf{k}$
\begin{align}
i\rightarrow \mathbf{k}=\mathbf{k}_i
\end{align}
For a system with $N$ unit cells, there exist a vast number of such one-to-one mappings. 
Here we can choose an arbitrary one of them, and the conclusions below are independent of this choice.
Utilizing this mapping that we choose, the wavefunction overlap between $\ket{\psi_I}$ and $\ket{\psi_{II}}$ can be factorized 
\begin{align}
|\phi|=|\braket{\psi_I|\psi_{II}}|=\prod_i |\phi_i|
\end{align}
where
\begin{align}
\phi_i=\langle 0| d_{\textrm{b},\mathbf{k}_i}& c_{\textrm{b},\mathbf{k}_i}  d_{t,\mathbf{k}_i}c_{\textrm{t},\mathbf{k}_i}
 (a_{\textrm{t},i,\uparrow}^\dagger a_{\textrm{b},i,\downarrow}^\dagger-a_{\textrm{t},i,\downarrow}^\dagger a_{\textrm{b},i,\uparrow}^\dagger)
\nonumber\\
&(b_{\textrm{t},i,\uparrow}^\dagger b_{\textrm{b},i,\downarrow}^\dagger-b_{\textrm{t},i,\downarrow}^\dagger b_{\textrm{b},i,\uparrow}^\dagger)
|0 \rangle
\end{align}
Here, for each $i$, this overlap only involves four creation (annihilation) operators, and thus $\phi_i$ doesn't suffer from the orthogonality catastrophe.
Because the two regimes (non-interacting and infinite-interaction) have ground states with different topology, we expect at least one $i$, at which $\phi_i$ vanishes. This is indeed the case for the model considered here.

\section{Discussion}
In this article, we explored the relation between wavefunction overlap and adiabatic continuity in (non-interacting) band insulators 
and interacting quantum systems. Our results can be utilized to simplify certain problems in the study of topological states. 
For example, in the study of band insulators, a large number of topological indices have been introduced 
(e.g. the Chern number, the Z$_2$ topological index, 
the mirror Chern number, the spin Chern number, the Hopf index), 
and more topological indices can be defined, if we enforce additional symmetries (e.g. space-group symmetries). 
As a result, to fully determine the topological property of an insulator becomes a nontrivial task. 
In principle, it is necessary to compute {\it all} these topological indices in order to achieve such an objective.
The conclusions reported in this article offer an alternative approach. Instead of trying to compute all known topological indices, 
one can utilize some known insulators as reference states, whose wavefunctions and topological properties are well understood. 
If the Bloch waves of a new insulator have nonzero overlap with some reference insulator, we immediately know 
the topological properties of this new insulator, which must be identical to the reference insulator. 
If the new insulator has zero Bloch-wavefunction overlap with all known reference insulators, then
this insulator might be a new topological state, and it requires further investigation to understand 
its topological structure.

It is worthwhile to notice that a nonzero wavefunction overlap is a sufficient condition for topologically equivalence, but it is not necessary. For example, two topologically equivalent states may accidentally have wavefunctions that are orthogonal to each other. Such an accidental vanishing wavefunction overlap is typically not stable and will be removed by small perturbations, while the topologically-protected zero wavefunction overlap is stable and cannot be removed.

For interacting systems, our theorem can be easily generalized. However, it cannot be applied to generic interacting systems 
because of the orthogonality catastrophe. On the other hand, in the study of interacting topological states, many numerical 
methods can only handle finite-size systems (e.g. exact diagonalization or density matrix renormalization group). 
There, our conclusions will not suffer from the orthogonality catastrophe, and thus could benefit some of the numerical investigations.

Above, we proved that if we have two insulators with different topology, there must exist (at least) one momentum point, 
at which the overlap of the wavefunction vanishes. The vanishing overlap has direct experimental implications, if we consider 
tunneling between these two insulators, i.e. the vanishing wavefunction overlap can prohibit tunneling between the two insulators
at certain momentum point. In Ref.~\onlinecite{Yang2013}, it is shown that this is indeed the case when one studies tunneling 
between Chern insulators and conventional insulators, and between time-reversal invariant topological insulators and conventional 
insulators. Our results suggest that similar physics could be generalized for more generic topological states.

\begin{acknowledgments}
The work was supported by the National Science Foundation, under grants PHY-1402971 at the University of Michigan.
\end{acknowledgments}

\appendix
\section{Insulators with more than one valence bands}
In this section, we present proofs for conclusions discussed in Sec.~\ref{sec:multi_bands}.

\subsection{anti-commutators for the $c$ and $d$ operators}
\label{sec:commutator_c_and_d}
We first prove Eqs.~\eqref{eq:commutator_c_and_d_1} and~\eqref{eq:commutator_c_and_d_2}. In general, creation operators $c_n^\dagger$
and $d_m^\dagger$ are connected by the following unitary transformation
\begin{align}
d_{m,\mathbf{k}}^\dagger=\sum_{n=1}^{+\infty} \braket{0|c_{n,\mathbf{k}} d_{m,\mathbf{k}}^\dagger|0} c_{n,\mathbf{k}}^\dagger
\label{eq:transformation_c_to_d}
\\
c_{n,\mathbf{k}}^\dagger=\sum_{m=1}^{+\infty} \braket{0|d_{m,\mathbf{k}} c_{n,\mathbf{k}}^\dagger|0} d_{m,\mathbf{k}}^\dagger
\label{eq:transformation_d_to_c}
\end{align}
We emphasize that in these two equations, the band indices $n$ and $m$ are summed over all bands (conduction and valence).
Utilizing Eq.~\eqref{eq:transformation_c_to_d},  it is straightforward to verify that
\begin{align}
\{c_{n,\mathbf{k}},d_{m,\mathbf{k}'}^\dagger\}
=\sum_{n'=1}^{+\infty} \braket{0|c_{n',\mathbf{k}'} d_{m,\mathbf{k}'}^\dagger|0}\{c_{n,\mathbf{k}}, c_{n',\mathbf{k}'}^\dagger\}
\nonumber\\
=\sum_{n'=1}^{+\infty} \mathcal{F}_{n',m} \delta_{n,n'}\delta_{\mathbf{k},\mathbf{k}'}
=\mathcal{F}_{n,m} \delta_{\mathbf{k},\mathbf{k}'}
\end{align}
Similarly, using Eq.~\eqref{eq:transformation_d_to_c}, we have
\begin{align}
\{d_{m,\mathbf{k}},c_{n,\mathbf{k}'}^\dagger\}=\sum_{m'=1}^{+\infty} \braket{0|d_{m',\mathbf{k}'} c_{n,\mathbf{k}'}^\dagger|0} 
\{d_{m,\mathbf{k}}, d_{m',\mathbf{k}'}^\dagger\}
\nonumber\\
=\sum_{m'=1}^{+\infty} \mathcal{F}^*_{n m'} \delta_{m,m'}\delta_{\mathbf{k},\mathbf{k}'}
= \mathcal{F}^*_{n,m}\delta_{\mathbf{k},\mathbf{k}'}
\end{align}

\subsection{the $\mathcal{F}$ and $\mathcal{U}$ matrices}
\label{sec:matrices}
In this section, we prove some properties of the $\mathcal{F}$ and $\mathcal{U}$ matrices.

\subsubsection{the existence of the $\mathcal{U}$ matrix}
First, we prove the existence of the $\mathcal{U}$ matrix. In the main text, we assumed that $\mathcal{U}$ is a unitary matrix, 
which diagonalizes the matrix $\mathcal{F}\mathcal{F}^\dagger$, i.e. $\mathcal{U}\mathcal{F}\mathcal{F}^\dagger\mathcal{U}^\dagger$ is a diagonal matrix.
To prove that such a $\mathcal{U}$ indeed exists, we just need to show that $\mathcal{F}\mathcal{F}^\dagger$ is a hermitian matrix, because
we know that any hermitian matrices can be diagonalized by some unitary matrices. Here, we compute directly the hermitian conjugate of 
$\mathcal{F}\mathcal{F}^\dagger$,
\begin{align}
(\mathcal{F}\mathcal{F}^\dagger)^\dagger=(\mathcal{F}^\dagger)^\dagger \mathcal{F}^\dagger=\mathcal{F}\mathcal{F}^\dagger
\end{align}
which indeed recovers itself, i.e. it is a hermitian matrix. As a result, there must exist some unitary matrix $\mathcal{U}$, such that
\begin{align}
\mathcal{U}_{l,n}  \mathcal{F}_{n,m} \mathcal{F}^*_{n',m}  \mathcal{U}^{*}_{l',n'}=\lambda_{l} \delta_{l,l'}
\label{eq:diagonalize_f_f_dagger}
\end{align}
where $\lambda_{l}$ are the eigenvalues of the matrix $\mathcal{F}\mathcal{F}^\dagger$. In this formula, we do not sum over $l$ for the right hand side (same below).

\subsubsection{$\lambda_{l}>0$}
Now, we will prove that the eigenvalues $\lambda_{l}$ are positive, as long as $\det \mathcal{F}\ne 0$, which will be used
later in Sec~\ref{sec:singular_free} when we prove that the normalization factor in the denominator 
is nonzero. We first prove that $\mathcal{F}\mathcal{F}^\dagger$
is semi-positive definite (i.e. all eigenvalues are non-negative), regardless of the value of $\det \mathcal{F}$. Then, we will further prove that
if $\det \mathcal{F}\ne 0$, the matrix $\mathcal{F}\mathcal{F}^\dagger$ is positive-definite (i.e.  all eigenvalues are are positive).

Assuming that $\mathbf{w}$ is an arbitrary row vector composed by  $N$ complex numbers, and $\mathbf{w}^\dagger$ is its 
hermitian conjugate. We know that
\begin{align}
\mathbf{w} \mathcal{F}\mathcal{F}^\dagger \mathbf{w}^\dagger= \mathbf{w} \mathcal{F} (\mathbf{w} \mathcal{F})^\dagger \ge 0
\end{align}
Because this result holds for any $\mathbf{w}$, $ \mathcal{F}\mathcal{F}^\dagger$ is semi-positive definite, i.e., its
eigenvalues are non-negative.

If $\det \mathcal{F}\ne 0$, $\det(\mathcal{F}\mathcal{F}^\dagger)=|\det \mathcal{F}|^2\ne 0$. Because the determinant of a hermitian matrix
equals to the product of all eigenvalues, this implies that none of the eigenvalues of the matrix $\mathcal{F}\mathcal{F}^\dagger$ is zero. Thus, this matrix is positive definite and all eigenvalues are positive.

\subsubsection{$\frac{\mathcal{F}^{*}_{n,m}\mathcal{U}^{*}_{l,n}\mathcal{U}_{l,n'}\mathcal{F}_{n',m'}}{\mathcal{N}_l^2}=\delta_{m,m'}$}
Now we prove that at $\alpha=1$, $\frac{\mathcal{F}^{*}_{n,m}\mathcal{U}^{*}_{l,n}\mathcal{U}_{l,n'}\mathcal{F}_{n',m'}}{\mathcal{N}_l^2}=\delta_{m,m'}$, which was utilzed to simplify 
Eq.~\eqref{eq:Hamiltonian_alpha_1_partial_simplified} in the main text.
First, we rewrite Eq.~\eqref{eq:diagonalize_f_f_dagger} in a matrix form
\begin{align}
\mathcal{U} \mathcal{F} \mathcal{F}^\dagger \mathcal{U}^\dagger=\mathcal{D}
\label{eq:matrix_form}
\end{align}
where $\mathcal{D}$ is a diagonal matrix 
\begin{align}
\mathcal{D}_{l,l'}=\lambda_{l} \delta_{l,l'}
\label{eq:dmatrix}
\end{align}
and $\lambda_l$ is the $l$th eigenvalue of the $ \mathcal{F} \mathcal{F}^\dagger$
matrix. We compute the matrix inverse for both sides of Eq.~\ref{eq:matrix_form}. Because $\mathcal{U}$ is a unitary matrix, $ \mathcal{U}^{-1}=\mathcal{U}^\dagger$, we have
\begin{align}
\mathcal{U} (\mathcal{F}^\dagger)^{-1} \mathcal{F}^{-1} \mathcal{U}^\dagger=\mathcal{D}^{-1}
\end{align}
And thus
\begin{align}
\mathcal{F}^\dagger\mathcal{U}^\dagger[\mathcal{U} (\mathcal{F}^\dagger)^{-1} \mathcal{F}^{-1} \mathcal{U}^\dagger] \mathcal{U}\mathcal{F}=\mathcal{F}^\dagger\mathcal{U}^\dagger\mathcal{D}^{-1}\mathcal{U}\mathcal{F}
\end{align}
If we simplify this equation, we find that
\begin{align}
\mathcal{I}=\mathcal{F}^\dagger\mathcal{U}^\dagger\mathcal{D}^{-1}\mathcal{U}\mathcal{F}
\end{align}
where $\mathcal{I}$ is the identity matrix.
If we write down the components for these matrices, we get
\begin{align}
\delta_{m,m'}=\mathcal{F}^*_{n,m}\mathcal{U}^*_{l,n} \mathcal{D}^{-1}_{l,l'} \mathcal{U}_{l',n'} \mathcal{F}_{n',m'}
\end{align}
Utilizing Eq.~\eqref{eq:dmatrix}, it is easy to realize that the inverse of the diagonal matrix $\mathcal{D}$ is
\begin{align}
\mathcal{D}^{-1}_{l,l'}=\lambda_{l}^{-1} \delta_{l,l'}
\end{align}
As shown in Eq.~\eqref{eq:normalization}, at $\alpha=1$, $\lambda_{l}^{-1}=1/\mathcal{N}_l^2$, and thus we have
\begin{align}
\delta_{m,m'}=\mathcal{F}^*_{n,m}\mathcal{U}^*_{l,n}\frac{\delta_{l,l'}}{\mathcal{N}_l^2} \mathcal{U}_{l',n'} \mathcal{F}_{n',m'}
=\frac{\mathcal{F}^{*}_{n,m}\mathcal{U}^{*}_{l,n}\mathcal{U}_{l,n'}\mathcal{F}_{n',m'}}{\mathcal{N}_l^2}.
\end{align}

\subsection{anti-commutators}
\label{sec:commutators_a}
Now, we compute the anti-commutators for $a$ and $a^\dagger$,
\begin{widetext}
\begin{align}
\{a_{l,\mathbf{k}}, a^\dagger_{l',\mathbf{k}'}\}
=&\frac{(1-\alpha)^2}{|\mathcal{N}_l|^2}  \mathcal{U}_{l,n}\mathcal{U}^{*}_{l',n'}\{c_{n,\mathbf{k}}, c^\dagger_{n',\mathbf{k}'}\}
+\frac{(1-\alpha)\alpha}{|\mathcal{N}_l|^2}  \mathcal{U}_{l,n}  \mathcal{U}^{*}_{l',n'}\mathcal{F}^*_{n',m'}\{c_{n,\mathbf{k}}, d^\dagger_{m',\mathbf{k}'}\}
\nonumber\\
&+\frac{(1-\alpha)\alpha}{|\mathcal{N}_l|^2}   \mathcal{U}_{l,n} \mathcal{F}_{n,m} \mathcal{U}^{*}_{l',n'}
\{d_{m,\mathbf{k}}, c^\dagger_{n',\mathbf{k}'}\}
+ \frac{\alpha^2}{|\mathcal{N}_l|^2}   \mathcal{U}_{l,n} \mathcal{F}_{n,m} \mathcal{U}^{*}_{l',n'}\mathcal{F}^*_{n',m'}\{d_{m,\mathbf{k}}, d^\dagger_{m',\mathbf{k}'}\}
\nonumber\\
=&\frac{(1-\alpha)^2}{|\mathcal{N}_l|^2}  \mathcal{U}_{l,n}\mathcal{U}^{*}_{l',n'} \delta_{n,n'}\delta_{\mathbf{k},\mathbf{k}'}
+\frac{(1-\alpha)\alpha}{|\mathcal{N}_l|^2}  \mathcal{U}_{l,n}  \mathcal{U}^{*}_{l',n'}\mathcal{F}^*_{n',m'}
\mathcal{F}_{n,m'}\delta_{\mathbf{k},\mathbf{k}'}
\nonumber\\
&+\frac{(1-\alpha)\alpha}{|\mathcal{N}_l|^2}   \mathcal{U}_{l,n} \mathcal{F}_{n,m} \mathcal{U}^{*}_{l',n'}
\mathcal{F}^{*}_{n',m}\delta_{\mathbf{k},\mathbf{k}'}
+ \frac{\alpha^2}{|\mathcal{N}_l|^2}  \mathcal{U}_{l,n} \mathcal{F}_{n,m} \mathcal{U}^{*}_{l',n'}\mathcal{F}^*_{n',m'} \delta_{m,m'}\delta_{\mathbf{k},\mathbf{k}'}
\nonumber\\
=&\frac{(1-\alpha)^2}{|\mathcal{N}_l|^2}  \mathcal{U}_{l,n}\mathcal{U}^{*}_{l',n} \delta_{\mathbf{k},\mathbf{k}'}
+\frac{\alpha(2-\alpha)}{|\mathcal{N}_l|^2}  \mathcal{U}_{l,n}  \mathcal{F}_{n,m} \mathcal{F}^*_{n',m}  \mathcal{U}^{*}_{l',n'}\delta_{\mathbf{k},\mathbf{k}'}
\end{align}
\end{widetext}
In the first term, because $\mathcal{U}$ is a unitary matrix, we have $\mathcal{U}_{l,n}\mathcal{U}^{*}_{l',n}=\delta_{l,l'}$. 
For the second term, we have shown in Eq.~\eqref{eq:diagonalize_f_f_dagger} that
$\mathcal{U}_{l,n}  \mathcal{F}_{n,m} \mathcal{F}^*_{n',m}  \mathcal{U}^{*}_{l',n'}=\lambda_{l} \delta_{l,l'}$,
where $\lambda_{l}$ is the $l$th eigenvalue of the matrix $\mathcal{F} \mathcal{F}^\dagger$.  As a result,
\begin{align}
\{a_{l,\mathbf{k}}, a^\dagger_{l',\mathbf{k}'}\}=\frac{(1-\alpha)^2+\alpha(2-\alpha) \lambda_{l}}{|\mathcal{N}_l|^2}\delta_{l,l'}
\delta_{\mathbf{k},\mathbf{k}'}
\label{eq:commutator_a_partial_result}
\end{align}
If we set the normalization factor 
\begin{align}
\mathcal{N}_l=\sqrt{(1-\alpha)^2+\alpha(2-\alpha) \lambda_{l}}
\label{eq:normalization}
\end{align}
the canonical anti-commutation relation is proved
\begin{align}
\{a_{l,\mathbf{k}}, a^\dagger_{l',\mathbf{k}'}\}=\delta_{l,l'}
\delta_{\mathbf{k},\mathbf{k}'}
\end{align}

\subsection{normalization factor}
\label{sec:singular_free}
In this section, we prove that the normalization factor defined in Eq.~\eqref{eq:normalization} never becomes zero. 
Because this normalization factor is used as a denominator in the definition of $a^\dagger_{l,\mathbf{k}}$, 
the fact that $\mathcal{N}_l\ne 0$ ensures that $a^\dagger_{l,\mathbf{k}}$ is not singular.


In Sec.~\ref{sec:matrices}, we have proved that as long as the overlap function is nonzero, $\lambda_{l}$ is positive.
For a positive $\lambda_{l}$ and $0\le\alpha\le 1$, it is easy to verify that $(1-\alpha)^2+\alpha(2-\alpha) \lambda_{l}>0$. 
Thus, according to Eq.~\eqref{eq:normalization}, we proved that $\mathcal{N}_l>0$.

\section{Symmetry of the adiabatic path}
\label{sec:symmetry}
In this section, we prove that for two quantum states with finite wavefunction overlap, the adiabatic path defined in the main text
preserves all the symmetries of the two quantum states.

\subsection{interacting systems}
We start by examining the symmetry of the adiabatic path defined in Eq.~\eqref{eq:Hamiltonian_interacting}.
Here, we consider unitary symmetries, but all the conclusions can be easily generalized to anti-unitary symmetries.
In quantum mechanics, a symmetry in a quantum state implies that the wavefunction must remain invariant under certain
transformation (e.g. translation, space inversion, etc) up to some possible $U(1)$ phase factor
\begin{align}
\ket{\psi}\rightarrow e^{i\varphi}\ket{\psi}
\\
\ket{\psi'}\rightarrow e^{i\varphi'}\ket{\psi'}
\end{align}
If these relations hold for $\ket{\psi}$ and $\ket{\psi'}$, it is straightforward to prove that under the same 
transformation, the wavefunction defined in Eq.~\eqref{eq:wavefunction} transforms as
\begin{align}
\ket{\Psi(\alpha)}\rightarrow  e^{i\varphi} \ket{\Psi(\alpha)}
\end{align} 
same as the state $\ket{\psi}$.
The corresponding bra vector transforms as 
\begin{align}
\bra{\Psi(\alpha)}\rightarrow  e^{-i\varphi} \bra{\Psi(\alpha)}
\end{align} 
where the complex phase takes the opposite sign.
As a result, the Hamiltonian defined in Eq.~\eqref{eq:Hamiltonian_interacting} is invariant under this 
transformation, because the phase factors from the bra and ket vectors cancel each other, i.e. the Hamiltonian 
preserves this symmetry.

\subsection{band insulators with one valence band}
Now we consider band insulators with one valence band, i.e. the Hamiltonian Eq.~\eqref{eq:Hamiltonian_one_band}.
Again,  we consider unitary symmetries, but all the conclusions can be easily generalized to anti-unitary symmetries.
Assume that insulators $I$ and $II$ preserve some symmetry.
Under the symmetry transformation, we assume that the momentum points are transformed as
\begin{align}
\mathbf{k} \rightarrow \mathbf{k}'.
\end{align}
and the Bloch waves are transformed according to certain unitary matrices. Because the insulator
is invariant under the transformation, this unitary matrix will {\it not} mix conduction and valence bands.
Since we have only one valence band, the Bloch waves of the valence band can only change by 
a phase shift under this symmetry transformation
\begin{align}
\ket{\psi^{I}(\mathbf{k})} \rightarrow e^{i \varphi(\mathbf{k})} \ket{\psi^{I}(\mathbf{k})}
\end{align}
Because the insulator is invariant under this transformation, we know that 
$e^{i \varphi(\mathbf{k})} \ket{\psi^{I}(\mathbf{k})}$ must be identical to the Bloch wave of the valence band at 
$\mathbf{k'}$
\begin{align}
\ket{\psi^{I}(\mathbf{k})} \rightarrow \ket{\psi^{I}(\mathbf{k}')}=e^{i \varphi(\mathbf{k})} \ket{\psi^{I}(\mathbf{k})}
\end{align}

For insulator $II$, the wavefunction satisfies the same relation, but the phase factor could be different
\begin{align}
\ket{\psi^{II}(\mathbf{k})} \rightarrow \ket{\psi^{II}(\mathbf{k}')} =e^{i \varphi'(\mathbf{k})} \ket{\psi^{II}(\mathbf{k})}.
\end{align}
As a result, the overlap function must satisfy
\begin{align}
\phi(\mathbf{k}')=e^{i[\varphi'(\mathbf{k})-\varphi(\mathbf{k})]} \phi(\mathbf{k})
\end{align}•
It is easy to verify that for the Bloch state $\ket{\Psi(\mathbf{k},\alpha)}$ defined in Eq.~\eqref{eq:bloch_one_band}, 
we have 
\begin{align}
\ket{\Psi(\mathbf{k},\alpha)}\rightarrow \ket{\Psi(\mathbf{k}',\alpha)}=e^{i \varphi(\mathbf{k})} \ket{\Psi(\mathbf{k},\alpha)}
\end{align}
same as $\ket{\psi^{I}}$. 
And thus the Hamiltonian that we defined for the adiabatic path [Eq.~\eqref{eq:Hamiltonian_one_band}] remains
invariant, i.e., it preserves the symmetry
\begin{align}
H(\alpha)\rightarrow H(\alpha) 
\end{align}

\subsection{band insulators with more than one valence bands}
In this section, we consider more generic band insulators with multiple valence bands.
Assume that insulators $I$ and $II$ preserve some unitary symmetry and 
under the symmetry transformation the momentum points transform as
\begin{align}
\mathbf{k} \rightarrow \mathbf{k}'.
\end{align}
Because insulator $I$ preserves the symmetry, under the symmetry transformation, Bloch waves
of the valence bands must satisfy,
\begin{align}
\ket{\psi^{I}_{n}(\mathbf{k})}\rightarrow
\ket{\psi^{I}_n(\mathbf{k}')} =\mathcal{U}^I_{n,n'} (\mathbf{k}) 
\ket{\psi^{I}_{n'}(\mathbf{k})}
\end{align}
where $ \mathcal{U}^I(\mathbf{k})$ is some unitary matrix that describe the transformation of the Bloch waves
under the symmetry transformation.
For insulator $II$, if the same symmetry is preserved, the wavefunction is transformed in a similar way, 
but the unitary matrix could be different
\begin{align}
\ket{\psi^{II}_{m}(\mathbf{k})}\rightarrow \ket{\psi^{II}_m(\mathbf{k}')}=\mathcal{U}^{II}_{m,m'}(\mathbf{k}) \ket{\psi^{II}_{m'}(\mathbf{k})}
\end{align}
As a result, the overlap matrix $\mathcal{F}_{n,m}=\braket{\psi^{I}_n|\psi^{II}_m}$ satisfies 
\begin{align}
\mathcal{F}_{\mathbf{k}'}=(\mathcal{U}^{I}_\mathbf{k})^{*}  \mathcal{F}_{\mathbf{k}} (\mathcal{U}_\mathbf{k}^{II})^{T} 
\label{eq:F_transformation}
\end{align}
where $*$ and $T$ stand for complex conjugate and transpose respectively. Here, we write the momentum as a subindex
to simplify the formula (same below).

As a result, we know that
\begin{align}
\mathcal{F}_{\mathbf{k}'}\mathcal{F}_{\mathbf{k}'}^\dagger =
(\mathcal{U}^{I}_\mathbf{k})^{*}  \mathcal{F}_{\mathbf{k}} \mathcal{F}_{\mathbf{k}}^\dagger   (\mathcal{U}^{I}_\mathbf{k})^{T} 
\end{align}
In the main text, we defined a $\mathcal{U}$ matrix at each momentum point to diagonalize the 
$\mathcal{F}\mathcal{F}^\dagger$ matrix. The relation above implies that
\begin{align}
\mathcal{U}_{\mathbf{k}'}=\mathcal{U}_{\mathbf{k}} (\mathcal{U}^{I}_{\mathbf{k}})^{T}
\label{eq:U_transformation}
\end{align}
up to some unimportant gauge choice (i.e. phase factors).

Utilizing Eqs.~\eqref{eq:F_transformation} and~\eqref{eq:U_transformation}, we can verify easily that
for the valence-band Bloch states defined in Eq~\eqref{eq:bloch_multi_band}, 
$\ket{\Psi(\mathbf{k},\alpha)}=\ket{\Psi(\mathbf{k}',\alpha)}$ up to a gauge choice. Thus, the insulator
that we defined as the adiabatic path preserves the correct symmetry.

\section{wavefunction overlap between different models}
\label{sec:differentmodel}
For the Kane-Mele model, we consider a honeycomb lattice with lattice constant set to unity. For
this lattice, the Hamiltonian can be written as
\begin{align}
H_{\text{KM}}&=d_1 I_{2\times 2} \otimes \tau_1+d_2 I_{2\times 2} \otimes \tau_2+d_3 \sigma_3 \otimes \tau_3.
\end{align}
where $I_{2\times 2}$ is the two-by-two identity matrix. $\sigma_i$ and $\tau_i$ represent Pauli matrices, and
\begin{align}
d_1&=-t_1 [1+\cos(k_1)+\cos(k_2)];\\
d_2&=-t_1 [\sin(k_1)+\sin(k_2)];\\
d_3&=-2 t_2 \sin(\phi) [\sin(k_1)-\sin(k_2)-\sin(k_1-k_2)];
\end{align}
where $k_1=\frac{1}{2}k_x+\frac{\sqrt{3}}{2}k_y$, $k_2=-\frac{1}{2}k_x+\frac{\sqrt{3}}{2} k_y$. $t_1$ and $t_2$ are the nearest-neighbor and next-nearest-neighbor hopping strengths, and 
$\phi$ is the phase change along with the next-nearest-neighbor hopping. 
The Brillouin zone for this model can be chosen as a rhombus formed by reciprocal lattice vectors $\frac{4\pi}{\sqrt{3}} (\frac{\sqrt{3}}{2}, \pm \frac{1}{2})$.

For the Bernevig-Hughes-Zhang model, we use the following Hamiltonian
\begin{align}
H_{\text{BHZ}}=&\sin(k_x) \sigma_3 \otimes \tau_1+ \sin(k_y) I_{2\times 2} \otimes \tau_2
\nonumber\\
&+[2-m-\cos(k_x)-\cos(k_y)] I_{2\times 2} \otimes \tau_3.
\end{align}
And the first Brillouin zone is a square region $(-\pi, \pi]\times (-\pi, \pi]$.

Notice that for both models, we choose the gauge such that the Hamiltonians remain invariant when we shift the momentum by a reciprocal lattice vector. For control parameters, we set $t_1=3$, $t_2=1$, $\phi=\pi/2$ for the Kane-Mele model and $m=1$ for the Bernevig-Hughes-Zhang model. When we fill the lowest two of the four bands, the ground states for both models share the same spin Chern number, i.e. the two insulators are topologically equivalent. In order to compare the wavefunctions for these two insulators, we map the Brillouin zone of the Kane-Mele model to that of the Bernevig-Hughes-Zhang model using the following mapping
\begin{align}
(k_x,k_y)\mapsto (-\pi +\frac{1}{2} k_x+\frac{\sqrt{3}}{2} k_y, -\pi +\frac{1}{2} k_x-\frac{\sqrt{3}}{2} k_y)
\end{align}
It is worthwhile to emphasize that there are other ways to map the Brillouin zone of one model to that of the other. As long as the mapping is a continuous bijection, 
it can be used to compute the wavefunction overlap.

With this mapping, we can now compute the wavefunction overlap for each momentum point, using the technique discussed above. The result for these two models is shown in Fig.~\ref{fig:KMandBHZ}, where the overlap remains finite and never reaches zero.

%
\end{document}